%% file: mnpaper.tex
\documentclass[usenatbib]{mn2e}

\usepackage{graphicx}
\usepackage{natbib}
\def\citeN{\citet}
\def\cite{\citep}

\footnotesize
\newdimen\digitwidth    
\setbox0=\hbox{\rm0}
\digitwidth=\wd0
\catcode`!=\active
\def!{\kern\digitwidth}
\normalsize
\title[A mildly recycled pulsar in an eccentric binary]{PSR J1753--2240: A mildly recycled pulsar in an eccentric binary system}
\author[M.~J.~Keith et al.]
{M.~J.~Keith$^{1,2}$\thanks{Email: mkeith@pulsarastronomy.net},
M.~Kramer$^1$,
A.~G.~Lyne$^1$,
R.~P.~Eatough$^1$,
I.~H.~Stairs$^3$,\newauthor
A.~Possenti$^4$,
F.~Camilo$^5$ and
R.~N.~Manchester$^2$
\\
$^1$ University of Manchester, Jodrell Bank Centre for Astrophysics, Alan Turing Building, Manchester M13 9PL, UK\\
$^2$ Australia Telescope National Facility, CSIRO, P.O. Box 76, Epping, NSW 1710, Australia\\
$^3$ Department of Physics and Astronomy, University of British Columbia,Vancouver, BC V6T 1Z1, Canada\\
$^4$ INAF - Osservatorio Astronomico di Cagliari, Poggio dei Pini, 09012 Capoterra, Italy\\
$^5$ Columbia Astrophysics Laboratory, Columbia University, New York, NY 10027, USA\\
}
\let\leqslant=\leq 
\date{}
\begin{document}

\maketitle
\newcommand{\setthebls}{
}

\setthebls

\begin{abstract} 
\input{abstract}

\end{abstract}


\input{paper_text_v12}

\bibliographystyle{mnras}
\bibliography{journals,myrefs,psrrefs,modrefs}

\end{document}

%% file: abstract.tex
We report the discovery of PSR J1753$-$2240 in the Parkes Multibeam
Pulsar Survey database.  This 95-ms pulsar is in an eccentric binary
system with a 13.6-day orbital period.  Period derivative measurements
imply a characteristic age in excess of 1 Gyr, suggesting that the
pulsar has undergone an episode of accretion-induced spin-up.  The
eccentricity and spin period are indicative of the companion being a
second neutron star, so that the system is similar to that of PSR
J1811$-$1736, although other companion types cannot be ruled out at
this time.  The companion mass is constrained by geometry to lie
above 0.48 solar masses, although long-term timing
observations will give additional constraints.  If the companion is a
white dwarf or main sequence star, optical observations may yield a
direct detection of the companion.  If the system is indeed one of the
few known double neutron star systems, it would lie significantly far
from the recently proposed spin-period/eccentricity relationship.

%% file: paper_text_v12.tex
\section{Introduction}

Several types of binary pulsar companion are known:
main sequence stars, neutron stars, white dwarfs and even
planetary-mass bodies \cite{ls05a}.  Pulsars in binary systems,
especially those with neutron star companions, are valuable tools for
studies of a variety of physics and astrophysics.  However, white dwarfs are the most common type of companion, representing
$\sim90\%$ of all pulsar companions.  White dwarf and neutron star
companions are usually part of a binary pulsar system in which the
pulsar has undergone a ``recycling'' process (e.g.~Alpar et
al.~1982)\nocite{acrs82}.  In this process, the pulsar forms first,
spins down and ceases its normal radio emission phase before the
companion evolves into a giant star, when it is likely to start
overflowing its Roche lobe.  Matter and angular momentum are
transferred to the pulsar in an accretion process during which the
system may be observable as an X-ray binary system \cite{bv91}.  In simple terms, the
duration of the accretion process depends to a large extent on the mass of the companion
star.  A low-mass companion will provide a long-lived accretion flow,
allowing the pulsar to spin up to a period of a few milliseconds and to reappear
as a radio source. In cases of more massive companions, a high-mass
X-ray binary is formed in this ``standard scenario''. Alternative
models involving a ``double core scenario'' have been proposed
\cite{bro95,dps06}.
Common to both models is the occurrence of an
evolutionary stage before the second supernova explosion (SN) which
involves a helium star and the neutron star in a tight circular orbit
\cite{vt84,dp03b}.  Depending on the orbital period and the mass of
the helium star, matter may be transferred from the helium star to the
neutron star (e.g. \citealp{dv04}), for a period of time which determines the final spin
period of the pulsar. The result is a mildly recycled pulsar with a
period of tens of milliseconds, being significantly greater than the
few milliseconds of a fully recycled pulsar.

Significant eccentricity is only expected if a binary system experiences a second supernova explosion, which may be the case in a high mass system.
If the system survives this further SN, a double neutron star
system (DNS) will be formed. Such DNSs are rare, only eight
being known in the Galactic disk. The most exciting DNSs are compact,
relativistic binaries in which a number of relativistic effects can be
observed, so that they can be used for tests of theories of
gravity. The best examples for this type are the Hulse-Taylor pulsar,
PSR B1913+16 \cite{ht75a} and the Double Pulsar, PSR J0737$-$3039A/B
\cite{bdp+03,lbk+04}. While in the Double Pulsar both neutron stars are
active and detectable radio pulsars, usually only the old, recycled
pulsar is observed. An exception is the case of PSR J1906+0746, where
we seem to observe the second-born, young pulsar \cite{lsf+06,kas08}. Other,
less compact types of DNS with orbital periods of many days rather
than hours exist, and it is likely that some of them have been 
formed via a wind-accretion phase rather than an accretion disk
process as in the standard scenario (e.g.~\citealp{dpp05}).

It seems likely that the formation history of the DNS, and in particular the
properties of the helium star, are imprinted onto the currently observed
system parameters. For instance, a correlation between the spin-period of the
pulsar and the system's eccentricity is observed
\cite{mlc+05,fkl+05}. Based upon seven DNSs known at the time, it was
argued that this relationship can be understood on evolutionary grounds
\cite{fkl+05} and the occurrence of low-velocity kicks during the second SN
explosion \cite{dpp05,van07}. In the latter scenario, the SN is
essentially symmetric and the post-SN properties are largely determined by the
amount of mass lost during the explosion, such that greater mass loss
leads to wider orbits and larger eccentricities.
This low-kick scenario is particularly interesting, as together with the
proposed formation of neutron stars in DNSs via an electron-capture SN
\cite{pdl+05}, it may explain simultaneously the spin-period eccentricity
relationship and the low masses of the second born neutron star in some recent
DNS discoveries \cite{van07}.
In contrast, in an asymmetric SN, a kick and a corresponding large velocity are
imparted to the neutron star and also determines the post-SN orbital
configuration. Evidence for large kicks is observed in
isolated pulsars \cite{hllk05} and can be inferred from the observations
of geodetic precession in the binary pulsars PSR B1913+16 \cite{kra98}
and PSR B1534+12 \cite{sta04} (see also review by \citealp{kvw08}).
Clearly, the discovery of more Galactic disk DNSs will
be extremely valuable to test and check the viability of the
correlation and the proposed models.

In this Paper we describe the discovery and subsequent timing observations of
PSR J1753$-$2240 which appears as a mildly-recycled 95-ms radio pulsar in an
eccentric 14-d orbit. We argue that the system parameters suggest that
the observable pulsar is the old component of a DNS, which allows us to
investigate the implication for the spin-period/eccentricity
relationship mentioned above.

\section{Discovery \& Follow-up timing}

\begin{figure}
\includegraphics[height=8cm,angle=-90]{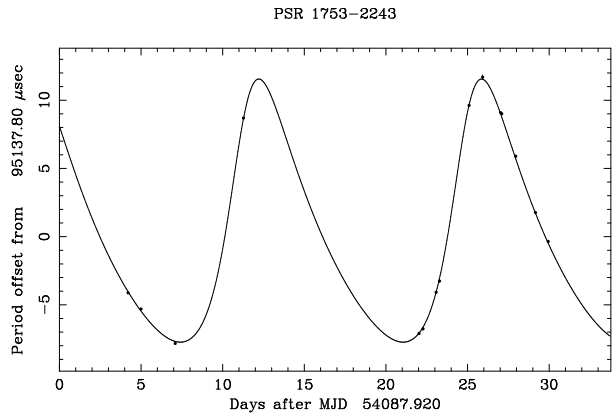}
\caption[Orbital period variation of PSR J1753$-$2240]{
\label{1753_orbit}
The apparent spin-period variation of PSR J1753$-$2240, showing the Doppler
effect due to binary motion. The points show the barycentric period
observed and the solid line shows the binary model with a
13.6-day orbital period and an eccentricity of 0.3 (Table~\ref{1753_params}). }
\end{figure}

\begin{figure}
\centering
\includegraphics[height=8cm,angle=-90]{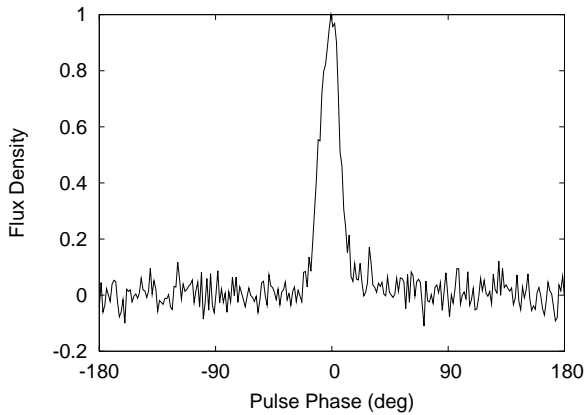}
\caption[Pulse profile of PSR J1753$-$2240]{
\label{1753_profile}
The integrated pulse profile of PSR J1753$-$2240 at 1374 MHz produced by summing data from 25 separate observations taken at Parkes.
There are 256 phase bins across the profile.
The y-axis is measured in arbitrary units of flux density.
}
\end{figure}

PSR J1753$-$2240 was discovered during the re-assessment of candidates from the
highly-successful Parkes Multibeam Pulsar Survey (e.g. \citealp{mlc+01}).  These
candidates were produced during the processing of these data, as described by
\citeN{fsk+04} and were re-evaluated using improved data-mining tools
developed and described by \citeN{kelk08}.  Following
the identification and selection of appropriate candidates,
observations were carried out at the Parkes telescope to
confirm the pulsar origin of the detected signals.  The receiver
system was centred on 1374~MHz and used the original 96-channel survey filterbank
and instrumentation (see Manchester et al.~2001 for details). Among
the 30 newly-discovered pulsars, the source designated PSR J1753$-$2240
showed significant variations in its 95-ms period between various
observations, indicating the presence of Doppler effects arising from
an orbit around a binary companion to the pulsar.  The apparent
spin-period variations can be fully described by a 13.6-day eccentric
Keplerian orbit, as shown in Figure \ref{1753_orbit}.
In order to
precisely measure the positional, spin and orbital parameters, the
pulsar was observed approximately twice a month for a period of 655
days using the Parkes telescope, producing 62 good time-of-arrival
(TOA) measurements. The standard pulsar timing procedure
(e.g.~\citealp{lk05}) was then applied to fit these TOAs for the
parameters using the {\sc tempo2} software, the results of which are
detailed in Table \ref{1753_params}.
The integrated pulse profile obtained from summing several of these observations is shown
in Figure~\ref{1753_profile}.

\section{Results \& Discussion}

The pulsar spin period of 95 ms is typical of that of a young pulsar,
although the period derivative is measured to be $\dot{P} =
(9.7 \pm 0.12) \times 10^{-19}$, implying a characteristic age of
greater than 1 Gyr and a surface magnetic field strength of 9.7$\times10^{9}$ G.  These parameters strongly suggest that the
pulsar's evolution included a period of recycling where the binary
companion has provided an episode of accretion induced spin-up.
However, the relatively long spin-period implies that the accretion
must have been relatively short-lived or otherwise ineffective.

\begin{table}
  \caption[Observed and derived parameters of PSR J1753$-$2240.]
   {Observed and derived parameters of PSR J1753$-$2240, after fitting
   to 62 TOAs measured from MJD 54028 to 54683.  The timing residuals
   relative to this model had an RMS of 0.4~ms.  Errors in
   the last digit are shown in parentheses}
\begin{center}
{\footnotesize
\begin{tabular}{ll}
\hline
\noalign{\smallskip}
 RA  (J2000 hh:mm:ss)          &  17:53:39.847(5)\\
 Dec (J2000 dd:mm:ss)          &  $-$22:40:42(1)\\
 Period (s)                 &  0.0951378086771(7)\\
 Period derivative (ss$^{-1}$)      &  0.00097(12) $\times10^{-15}$\\
 Epoch (MJD)                &  54275.02\\
 Solar System Ephemeris     &  DE405\\
 Dispersion Measure, DM (${\rm pc}$ ${\rm cm}^{-3}$) & 158.6(4)\\
 Flux Density at 1374~MHz (mJy)  & $0.15(3)$\\
 Pulse width at 10\% of peak (ms) & 8.6(6) \\ 
\noalign{\medskip}
Binary Parameters & BT model \\
 Orbital period (d)       & 13.6375668(7) \\
 Epoch of periastron passage (MJD)&  54099.13029(8)\\
 Projected semi-major axis (lt-s) & 18.11537(10)\\
 Longitude of periastron (deg)   & $-$49.3151(20)\\
 Eccentricity               & 0.303582(10) \\
\noalign{\medskip}
Characteristic Age (yr)   &1.6 $\times10^{9}$\\
Characteristic Magnetic Field (G) & 9.7 $\times10^{9}$\\
Mass function (M$_{\odot}$)&0.0343\\
Minimum Companion Mass (M$_{\odot}$)&0.4875\\
Distance$^*$ (kpc) & 3\\
\noalign{\smallskip}
\hline
\end{tabular}
{
\footnotesize $^*$ derived from the NE2001 model of Cordes \& Lazio (2002).}
}
\end{center}
\label{1753_params}
\end{table}
\nocite{cl02}

\subsection{Nature of the companion}

Unlike superficially similar systems in globular clusters such as
PSR J1748$-$2246J \cite{rhs+05}, the chance of creating binaries through
exchange interactions in the Galactic disk is believed to be
negligible. While recently an unusual and puzzling combination of
orbital parameters has been found in the case of PSR J1903$+$0327
\cite{crl+08}, the parameters observed here are commonly associated
with mildly recycled pulsars where the companion object was
responsible for the partial spin-up of the pulsar. Given the short
spin-period and high orbital eccentricity, we consider it therefore as
most likely that PSR J1753$-$2240 is a member of a DNS, with the
detected pulsar being the first-born pulsar.
In fact inspection of Table \ref{1811comp} clearly confirms that all the parameters of PSR J1753$-$2240 are in the typical range of DNS systems.
The only exception is the minimum companion mass, which is smaller than for the other known DNSs. However we note that the observed mass function
of the system allows for a companion significantly more massive than the minimum value shown.

Orbital parameters can provide some restrictions on
the companion mass, based upon geometry. Assuming a random distribution of
orbital orientation and a nominal pulsar mass of $1.4 M_{\odot}$ we
derive from the mass function at a $95\%$ confidence level that the
companion mass is less than $2.5 M_\odot$ (for $i \approx
18^\circ$). This implies a maximum system mass of around $4 M_{\odot}$
and a semi-major axis of approximately 36 light seconds.  If we allow
a reasonable range of pulsar and companion masses such that they lie between
1.2 and 1.8 $M_\odot$, the
measured mass function implies an inclination angle range of $22^\circ\le i
\le 34^\circ$, which has a nominal probability of $10\%$ of occurring
in a randomly selected system.

If the companion is a second neutron star, it is possible that it also
may be detectable as a radio pulsar.  Long, 2.5-hour observations of
PSR J1753$-$2240 using the same Parkes observing system at 1374 MHz
were carried out in a search for the companion, but no pulses have
been detected. The sensitivity of the search implies that the
companion, has an apparent 1400-MHz luminosity of less
than 0.9 mJy kpc$^2$. Only $\sim$6\% of all known pulsars with published
1400-MHz flux density measurements have luminosities below this limit.

Searches for a white-dwarf or a main-sequence companion in archival
optical and infra-red data sets have not been successful, although deeper
observations would be required to rule out either type of companion.
Given the DM estimated distance of $\sim 3$ kpc and the small Galactic
latitude of $b=1.5$ deg, a white dwarf would have a V-band magnitude
of $\sim 26$, which is within the reach of large optical telescopes.

\subsection{Spin-period -- eccentricity relationship}

If J1753$-$2240 is indeed a partially recycled pulsar in a DNS system, we can
consider how it relates to other DNSs. We are particularly interested
in how it fits into the relationship found between the orbital eccentricity
and the spin-period of the first-born pulsar (see Section 1). For this reason,
we exclude the two known DNSs J1906+0746 and B2127+11C. The latter is in
a globular cluster and was probably formed via an exchange interaction rather
than by binary evolution \cite{pakw91}. In contrast, PSR J1906+0746 is likely
to be a second-born young pulsar \cite{lsf+06} and the period of the
first-born neutron star is unknown. A comparison of the parameters
for the remaining seven DNS systems and J1753$-$2240 is given in Table
\ref{1811comp}.

\begin{figure}\centering
\includegraphics[width=7.5cm]{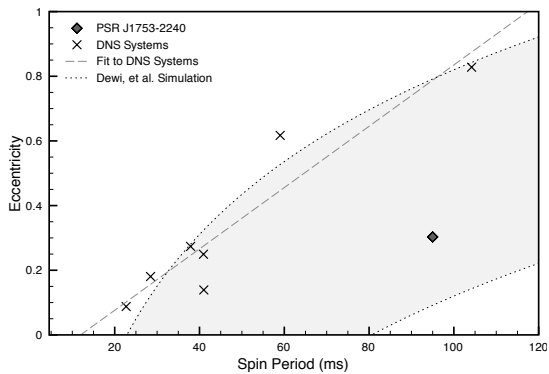}
\caption[Spin Period - Orbital Eccentricity Relationship]{
\label{1753_p_ecc}
The spin-period and eccentricity for the seven double neutron star
systems and PSR J1753$-$2240.  The long dashed line shows a linear fit
to the seven double neutron star systems.  The grey shaded area shows
the approximate region for which the simulations in
\citeN{dpp05} produced most systems. This area does not
reflect the density of simulated points in any way.
}
\end{figure}

In Figure~\ref{1753_p_ecc} we show the updated plot of eccentricity
versus orbital period as presented by \citeN{fkl+05} but now including
PSR J1753$-$2240. Our suggested DNS, PSR J1753$-$2240, has a much
smaller eccentricity or longer period than suggested by the correlation seen in the
other seven DNSs. 
Given the assumed hypothesis regarding the nature of the system there are three possible explanations for this:
(a) the system was
formed in an unlikely or unusual formation process with a relatively
small mass loss, or a rather fortuitous kick during an asymmetric
supernova explosion, (b) the previously inferred relationship between
spin-period and eccentricity arose from the low-number statistics,
or (c) the pulsar has spun down during its lifetime to move away from the
spin-eccentricity line. We consider these possibilities in turn.

It is possible that the formation mechanism of the J1753--2240 system
is somehow different from that of the other selected DNSs.
However, looking at Table \ref{1811comp}, there is little evidence to suggest that the J1753--2240
system is significantly different in nature from the other DNSs, we will assume
for the rest of this discussion that the formation mechanism of this system is not
unique amongst the known DNSs.

It is obviously very difficult to make a clear statement about the
reality of the spin period -- eccentricity relationship as the number
of involved DNS systems is still small. However, the possible physical
origin of such a relationship was also studied by \citeN{wakb08} who
investigated whether the observed relation can be produced by
assuming that kicks are restricted to be along the rotation axis of
the NS progenitor. Such an assumption is motivated by the observation
of an alignment of the velocities vectors of pulsars with their
rotation axis (Johnston et al.~2005)\nocite{jhv+05}.
It is not clear however that the kick must be aligned with the pre-SN
rotation axis as a post-SN alignment can also be produced for kicks directed
away from the pre-SN rotation axis, depending on the relative duration of the kick
mechanism and the initial spin period of the pulsar \cite{sp98}.

Assuming ``polar kicks'', i.e.~kicks to the
second-born neutron stars constrained to be within $10^\circ$ of the
progenitor's rotation axis, no correlation appears for large kick
velocities typical of isolated pulsars\cite{wakb08}. When low kicks are
imparted, the correlation can be well produced, supporting the case
for a low-kick scenario, at least in the case of the standard formation
scenario.
However, when constrained to a low kick velocity, the standard formation
models are unable to reproduce the spin-orbit mis-alignment angles observed in PSR B1913+16~\cite{wkh04}.
Hence, Willems et al. suggest that
systems like PSR B1913+16 may be formed via a second, different
formation channel, where no mass transfer occurs after the common
envelope phase and large kicks with magnitudes inferred for isolated
pulsars are imparted on the second-born neutron star. PSR J1753$-$2240
may represent such a system where the kick was large enough to produce
a significant eccentricity while the total amount of recycling was
small.

On the other hand, in the work of \citeN{dpp05,dpp07} no second formation
channel is invoked, but their simulations can also reproduce the observed
correlation by simply assuming that the births of the second born neutron stars
in DNSs are accompanied by moderate supernova kicks (less than 50 km s$^{-1}$)
and that the physical parameters of the DNSs are determined by the evolution
of Helium-star -- neutron-star binaries.  The orbital eccentricity is produced
primarily by the sudden mass loss in the second supernova explosion. Despite the
success in explaining the observed trend, there is a large spread in the final
properties of the simulated population, and the authors are forced to disregard
the kinematicaly derived large kicks for PSRs B1913+16 and B1534+12~\cite{wkh04}.
Inspecting Dewi et al.'s result, we
notice that the seven previously known DNSs are essentially lying along the
upper envelope of systems in the Dewi et al. simulations (see
Figure~\ref{1753_p_ecc}). At the same time, our newly discovered pulsar
appears much closer to the majority of the simulated pulsars than the known pulsars.  From
this point of view, it is suggestive to assume that the observed system
parameters are in fact typical and that the deviation of the new point from
the correlation simply reflects the statistical variation and the
different conditions that lead to the spread in system parameters seen by Dewi
et al.

The simulations by Dewi et al.~take the evolution of the orbital
parameters due to the emission of gravitational waves into account,
but it is unclear to us as to whether they also include the spin
evolution of pulsars. Due the small spin-down rates, the evolutionary
timescales are long, but one can speculate as to whether including the
spin evolution would tend to smear out any relationship existing at
birth of the systems.  Using the standard spin-period evolution model (e.g. \citealp{ls05a}), we can
estimate the time it may have taken the pulsar to move away from the
suggested line of the spin period-eccentricity relationship. For a
braking index of $n=3$, it would have taken $1.1\times 10^9$yr for the
pulsar to move from $P_0=40$ms to the observed period of $P=95$ms. As
the simulations by Dewi et al. in fact suggest a broad band for the
actual relationship, the actual evolution time could be even smaller,
but it raises the question why the previous known systems would then
still follow a common relationship. We suggest that future studies
consider this possible complication in some detail.

\begin{table*}
\caption[Outline of the parameters for 7 of the known DNS systems and J1753$-$2240.]{
	\label{1811comp} 
Summary of the parameters of the seven known
first-born pulsars in Galactic DNS systems and PSR~J1753$-$2240.  Pulsars
are ordered by their spin-period.  Except where shown in parentheses,
errors in values are smaller than the number of significant figures
shown.

}

{
	\begin{tabular}{l|llllllll}
	& J0737$-$3039A & J1756$-$2251 &B1534+12 & J1518+4904 & J1829+2456 & B1913+16 & J1753$-$2240 & J1811$-$1736 \\
		\hline
		$P_{\rm spin}$ (ms)             &\tt 22.69&\tt 28.46&\tt 37.90&\tt 40.93&\tt 41.01&\tt 59.03&\tt 95.13    & \tt 104.18 \\
		$\dot{P}_{\rm spin}$ ($\times 10^{-18}$)     &\tt 0.2 &\tt 0.1 &\tt 2  &\tt 0.03&\tt 0.05&\tt 8  &\tt 1& \tt 0.9  \\
		\\
		$P_{\rm binary}$  (day)         &\tt 0.10   &\tt 0.32   &\tt 0.42   &\tt 8.63   &\tt 1.17   &\tt 0.32   &\tt 13.63      & \tt 18.77   \\
		$A_p\sin{(i)}$  (lt-s)        &\tt 1.41   &\tt 2.75   &\tt 3.73   &\tt 20.04  &\tt 7.23   &\tt 2.34   &\tt 18.11      & \tt 34.78  \\
		$e$                         &\tt 0.08   &\tt 0.18   &\tt 0.27   &\tt 0.24   &\tt 0.14   &\tt 0.62   &\tt 0.30       & \tt 0.83   \\
		\\
		$\tau_c$ (MYr)               &\tt 204    &\tt 443    &\tt 249    &\tt 23000  &\tt 12400  &\tt 108    &\tt 1500  & \tt 1830    \\
		$B_{\rm surf}$ ($10^{10}$ G)     &\tt 6.4    &\tt 5.4    &\tt 9.7    &\tt 1.0    &\tt 1.5    &\tt 2.3    &\tt  0.97      & \tt 9.8    \\
		Min$[m_c]$ ($M_\odot$)       &\tt 1.25   &\tt 1.10   &\tt 1.30   &\tt 0.81   &\tt 1.26   &\tt 0.87   &\tt 0.49       & \tt 0.93  \\
		$m_c$                        &\tt 1.2489(7) &\tt 1.26(2)&\tt 1.3452(10)& -   & -      &\tt 1.3867(2)& -          & -      \\
		\\
		Reference                          &(1)&(2)&(3)&(4)&(5)&(6)&-&(7)\\

		\hline
		\end{tabular}
		References: (1)\citeN{ksm+06}, (2)\citeN{fer08}, (3)\citeN{sttw02}, (4)\citeN{jsk+08}, (5)\citeN{clm+05}, (6)\citeN{wt03}, (7)\citeN{cks+07}.

}
\end{table*}

\subsection{Future observations}

It has been argued \cite{pdl+05,van07} that low-kick births of neutron stars in DNS
systems may be the result of a new formation channel for neutron stars.  In
such a case, the neutron star may not be formed in the collapse of the
massive iron core, but instead in an ``electron-capture'' supernova, in which
a slightly less massive O-Ne-Mg core starts capturing electrons onto Mg to
initiate the collapse (Nomoto 1984, Podsiadlowski et al.~2004).
\nocite{nom84,plp+04} If such a process were indeed also responsible for
forming the observed system of PSR J1753$-$2240, then very definite
predictions could be made about the mass of the pulsar's companion,
since possible progenitor baryonic masses are expected in a small
range near $1.37\,M_{\odot}$.  The mass of the finally formed
companion neutron star would then have this progenitor mass less the
gravitational binding energy of the neutron star, suggesting companion
masses of around $1.25\,M_\odot$ (depending on the neutron star's
equation of state, Podsiadlowski et al.~2005). Further timing of PSR
J1753$-$2240 will eventually be able to test this hypothesis as it
will lead to better constraints on the total system mass, using a
future detection of a relativistic periastron advance. Simulations
using the {\sc tempo2} `fake' plugin \cite{hem06} suggest that given a
pulsar and companion mass of $\sim 1.25 M_\odot$ the advance of
periastron will be detectable at the 5-$\sigma$ level within 4
years. At the same time, we will conduct deep optical
observations in order to attempt detection of the optical emission of
any white dwarf or main-sequence star companion.

\section{Conclusions \& Summary}

PSR J1753$-$2240 shows many of the features of a DNS binary, however
it also shows some dissimilarities.  In the first instance, the
mass function of the system is somewhat lower than those of the other
similar DNS systems, however the required orbital inclination is not
unreasonable.  The system also lies far from the relationship between
spin period and eccentricity suggested for DNS systems. Nevertheless
we argue that the most likely companion is another neutron star,
principally because of the partially recycled nature of the system and
the observed orbital parameters.  Regardless of the companion type,
PSR J1753$-$2240 is an interesting system, lying in a poorly sampled
region of the spin period/orbital period/eccentricity phase space.
Therefore it will be important to continue observations, to further
constrain the system mass and also to monitor the long-term evolution
of the system.  If the DNS nature of the system is indeed confirmed
(or rejected) by future radio, optical or infra-red observations, the
system provides an interesting case for the question about high- or
low-kick formation of neutrons stars, illuminating the question about
the validity, range and/or origin of the spin-eccentricity
relationship.

\section*{Acknowledgements}
This research was partly funded by grants from the Science \& Technology Facilities Council, UK. The Australia Telescope
is funded by the Commonwealth of Australia for operation as a National
Facility managed by the CSIRO.
The authors would like to thank the following people for their contributions to observing at Parkes:
M. Burgay, A. Notsos, G.Hobbs, J. O'Brien, A. Corongiu, M. Purver, R. Smits and C. Espinoza.

%% file: mnpaper.bbl
\begin{thebibliography}{}

\bibitem[\protect\citeauthoryear{Alpar et~al.}{Alpar et~al.}{1982}]{acrs82}
Alpar M.~A., Cheng A.~F., Ruderman M.~A.,  Shaham J., 1982, Nature, 300, 728

\bibitem[\protect\citeauthoryear{Bhattacharya \& {van den Heuvel}}{Bhattacharya
  \& {van den Heuvel}}{1991}]{bv91}
Bhattacharya D.,  {van den Heuvel} E.~P.~J., 1991, Phys. Rep., 203, 1

\bibitem[\protect\citeauthoryear{Brown}{Brown}{1995}]{bro95}
Brown G.~E., 1995, ApJ, 440, 270

\bibitem[\protect\citeauthoryear{{Burgay} et~al.}{{Burgay}
  et~al.}{2003}]{bdp+03}
{Burgay} M. et~al., 2003, Nature, 426, 531

\bibitem[\protect\citeauthoryear{{Champion} et~al.}{{Champion}
  et~al.}{2005}]{clm+05}
{Champion} D.~J. et~al., 2005, MNRAS, 363, 929

\bibitem[\protect\citeauthoryear{{Champion} et~al.}{{Champion}
  et~al.}{2008}]{crl+08}
{Champion} D.~J. et~al., 2008, Science, 320, 1309

\bibitem[\protect\citeauthoryear{{Cordes} \& {Lazio}}{{Cordes} \&
  {Lazio}}{2002}]{cl02}
{Cordes} J.~M.,  {Lazio} T.~J.~W., preprint (arXiv:astro-ph/0207156)

\bibitem[\protect\citeauthoryear{{Corongiu} et~al.}{{Corongiu}
  et~al.}{2007}]{cks+07}
{Corongiu} A., {Kramer} M., {Stappers} B.~W., {Lyne} A.~G., {Jessner} A.,
  {Possenti} A., {D'Amico} N.,  {L{\"o}hmer} O., 2007, A\&A, 462, 703

\bibitem[\protect\citeauthoryear{{Dewi}, {Podsiadlowski}, \& {Pols}}{{Dewi}
  et~al.}{2005}]{dpp05}
{Dewi} J.~D.~M., {Podsiadlowski} P.,  {Pols} O.~R., 2005, MNRAS, 363, L71

\bibitem[\protect\citeauthoryear{{Dewi}, {Podsiadlowski}, \& {Pols}}{{Dewi}
  et~al.}{2007}]{dpp07}
{Dewi} J.~D.~M., {Podsiadlowski} P.,  {Pols} O.~R., 2007, in American Institute
  of Physics Conference Series, Vol. 924, {di Salvo} T., {Israel} G.~L.,
  {Piersant} L., {Burderi} L., {Matt} G., {Tornambe} A.,  {Menna} M.~T., ed,
  The Multicolored Landscape of Compact Objects and Their Explosive Origins, p.
  656

\bibitem[\protect\citeauthoryear{{Dewi}, {Podsiadlowski}, \& {Sena}}{{Dewi}
  et~al.}{2006}]{dps06}
{Dewi} J.~D.~M., {Podsiadlowski} P.,  {Sena} A., 2006, MNRAS, 368, 1742

\bibitem[\protect\citeauthoryear{{Dewi} \& {Pols}}{{Dewi} \&
  {Pols}}{2003}]{dp03b}
{Dewi} J.~D.~M.,  {Pols} O.~R., 2003, MNRAS, 344, 629

\bibitem[\protect\citeauthoryear{Dewi \& van~den Heuvel}{Dewi \& van~den
  Heuvel}{2004}]{dv04}
Dewi J.~D.~M.,  van~den Heuvel E.~P.~J., 2004, MNRAS, 349, 169

\bibitem[\protect\citeauthoryear{{Faulkner} et~al.}{{Faulkner}
  et~al.}{2005}]{fkl+05}
{Faulkner} A.~J. et~al., 2005, ApJ, 618, L119

\bibitem[\protect\citeauthoryear{{Faulkner} et~al.}{{Faulkner}
  et~al.}{2004}]{fsk+04}
{Faulkner} A.~J. et~al., 2004, MNRAS, 355, 147

\bibitem[\protect\citeauthoryear{Ferdman}{Ferdman}{2008}]{fer08}
Ferdman R.~D., 2008, Ph.D. thesis, Univ. British Columbia

\bibitem[\protect\citeauthoryear{Hobbs et~al.}{Hobbs et~al.}{2005}]{hllk05}
Hobbs G., Lorimer D.~R., Lyne A.~G.,  Kramer M., 2005, MNRAS, 360, 974

\bibitem[\protect\citeauthoryear{{Hobbs}, {Edwards}, \& {Manchester}}{{Hobbs}
  et~al.}{2006}]{hem06}
{Hobbs} G.~B., {Edwards} R.~T.,  {Manchester} R.~N., 2006, MNRAS, 369, 655

\bibitem[\protect\citeauthoryear{Hulse \& Taylor}{Hulse \&
  Taylor}{1975}]{ht75a}
Hulse R.~A.,  Taylor J.~H., 1975, ApJ, 195, L51

\bibitem[\protect\citeauthoryear{{Janssen} et~al.}{{Janssen}
  et~al.}{2008}]{jsk+08}
{Janssen} G.~H., {Stappers} B.~W., {Kramer} M., {Nice} D.~J., {Jessner} A.,
  {Cognard} I.,  {Purver} M.~B., 2008, A\&A, {in press (arXiv:0808.2292)}

\bibitem[\protect\citeauthoryear{{Johnston} et~al.}{{Johnston}
  et~al.}{2005}]{jhv+05}
{Johnston} S., {Hobbs} G., {Vigeland} S., {Kramer} M., {Weisberg} J.~M.,
  {Lyne} A.~G., 2005, MNRAS, 364, 1397

\bibitem[\protect\citeauthoryear{{Kalogera}, {Valsecchi}, \&
  {Willems}}{{Kalogera} et~al.}{2008}]{kvw08}
{Kalogera} V., {Valsecchi} F.,  {Willems} B., 2008, in American Institute of
  Physics Conference Series, Vol. 983, {Bassa} C., {Wang} Z., {Cumming} A.,
  {Kaspi} V.~M., ed, 40 Years of Pulsars: Millisecond Pulsars, Magnetars and
  More, p. 433

\bibitem[\protect\citeauthoryear{{Kasian}}{{Kasian}}{2008}]{kas08}
{Kasian} L., 2008, in American Institute of Physics Conference Series, Vol.
  983, {Bassa} C., {Wang} Z., {Cumming} A.,  {Kaspi} V.~M., ed, 40 Years of
  Pulsars: Millisecond Pulsars, Magnetars and More, p. 485

\bibitem[\protect\citeauthoryear{{Keith} et~al.}{{Keith} et~al.}{2008}]{kelk08}
{Keith} M.~J., {Eatough} R.~P., {Lyne} A.~G.,  {Kramer} M., 2008, MNRAS,
  {submitted}

\bibitem[\protect\citeauthoryear{{Kramer}}{{Kramer}}{1998}]{kra98}
{Kramer} M., 1998, ApJ, 509, 856

\bibitem[\protect\citeauthoryear{{Kramer} et~al.}{{Kramer}
  et~al.}{2006}]{ksm+06}
{Kramer} M. et~al., 2006, Science, 314, 97

\bibitem[\protect\citeauthoryear{Lorimer \& Kramer}{Lorimer \&
  Kramer}{2005}]{lk05}
Lorimer D.~R.,  Kramer M., 2005, Handbook of Pulsar Astronomy.
\newblock Cambridge University Press

\bibitem[\protect\citeauthoryear{{Lorimer} et~al.}{{Lorimer}
  et~al.}{2006}]{lsf+06}
{Lorimer} D.~R. et~al., 2006, ApJ, 640, 428

\bibitem[\protect\citeauthoryear{Lyne et~al.}{Lyne et~al.}{2004}]{lbk+04}
Lyne A.~G. et~al., 2004, Science, 303, 1153

\bibitem[\protect\citeauthoryear{Lyne \& Smith}{Lyne \& Smith}{2005}]{ls05a}
Lyne A.~G.,  Smith F.~G., 2005, Pulsar Astronomy, 3rd ed.
\newblock Cambridge University Press, Cambridge

\bibitem[\protect\citeauthoryear{Manchester et~al.}{Manchester
  et~al.}{2001}]{mlc+01}
Manchester R.~N. et~al., 2001, MNRAS, 328, 17

\bibitem[\protect\citeauthoryear{McLaughlin et~al.}{McLaughlin
  et~al.}{2005}]{mlc+05}
McLaughlin M.~A., Lorimer D.~R., Champion D.~J., Arzoumanian Z., Backer D.~C.,
  Cordes J.~M.,  Xilouris K.~M., 2005, in Astronomical Society of the Pacific
  Conference Series, Vol. 328, {Rasio} F.~A.,  {Stairs} I.~H., ed, Binary Radio
  Pulsars, p.~43

\bibitem[\protect\citeauthoryear{{Nomoto}}{{Nomoto}}{1984}]{nom84}
{Nomoto} K., 1984, ApJ, 277, 791

\bibitem[\protect\citeauthoryear{{Podsiadlowski} et~al.}{{Podsiadlowski}
  et~al.}{2005}]{pdl+05}
{Podsiadlowski} P., {Dewi} J.~D.~M., {Lesaffre} P., {Miller} J.~C., {Newton}
  W.~G.,  {Stone} J.~R., 2005, MNRAS, 361, 1243

\bibitem[\protect\citeauthoryear{{Podsiadlowski} et~al.}{{Podsiadlowski}
  et~al.}{2004}]{plp+04}
{Podsiadlowski} P., {Langer} N., {Poelarends} A.~J.~T., {Rappaport} S., {Heger}
  A.,  {Pfahl} E., 2004, ApJ, 612, 1044

\bibitem[\protect\citeauthoryear{Prince et~al.}{Prince et~al.}{1991}]{pakw91}
Prince T.~A., Anderson S.~B., Kulkarni S.~R.,  Wolszczan W., 1991, ApJ, 374,
  L41

\bibitem[\protect\citeauthoryear{{Ransom} et~al.}{{Ransom}
  et~al.}{2005}]{rhs+05}
{Ransom} S.~M., {Hessels} J.~W.~T., {Stairs} I.~H., {Freire} P.~C.~C., {Camilo}
  F., {Kaspi} V.~M.,  {Kaplan} D.~L., 2005, Science, 307, 892

\bibitem[\protect\citeauthoryear{Spruit \& Phinney}{Spruit \&
  Phinney}{1998}]{sp98}
Spruit H.,  Phinney E.~S., 1998, Nature, 393, 139

\bibitem[\protect\citeauthoryear{{Stairs}, {Thorsett}, \&
  {Arzoumanian}}{{Stairs} et~al.}{2004}]{sta04}
{Stairs} I.~H., {Thorsett} S.~E.,  {Arzoumanian} Z., 2004, Phys. Rev. Lett.,
  93, 141101

\bibitem[\protect\citeauthoryear{Stairs et~al.}{Stairs et~al.}{2002}]{sttw02}
Stairs I.~H., Thorsett S.~E., Taylor J.~H.,  Wolszczan A., 2002, ApJ, 581, 501

\bibitem[\protect\citeauthoryear{{van den Heuvel}}{{van den
  Heuvel}}{2007}]{van07}
{van den Heuvel} E.~P.~J., 2007, in American Institute of Physics Conference
  Series, Vol. 924, American Institute of Physics Conference Series, p. 598

\bibitem[\protect\citeauthoryear{van~den Heuvel \& Taam}{van~den Heuvel \&
  Taam}{1984}]{vt84}
van~den Heuvel E.~P.~J.,  Taam R.~E., 1984, Nature, 309, 235

\bibitem[\protect\citeauthoryear{{Weisberg} \& {Taylor}}{{Weisberg} \&
  {Taylor}}{2005}]{wt03}
{Weisberg} J.~M.,  {Taylor} J.~H., 2005, in Astronomical Society of the Pacific
  Conference Series, Vol. 328, {Rasio} F.~A.,  {Stairs} I.~H., ed, Binary Radio
  Pulsars, p.~25

\bibitem[\protect\citeauthoryear{{Willems} et~al.}{{Willems}
  et~al.}{2008}]{wakb08}
{Willems} B., {Andrews} J., {Kalogera} V.,  {Belczynski} K., 2008, in American
  Institute of Physics Conference Series, Vol. 983, {Bassa} C., {Wang} Z.,
  {Cumming} A.,  {Kaspi} V.~M., ed, 40 Years of Pulsars: Millisecond Pulsars,
  Magnetars and More, p. 464

\bibitem[\protect\citeauthoryear{Willems, Kalogera, \& Henninger}{Willems
  et~al.}{2004}]{wkh04}
Willems B., Kalogera V.,  Henninger M., 2004, ApJ, 616, 414

\end{thebibliography}
